\documentclass[aps,prl,twocolumn,nofootinbib]{revtex4}
\usepackage{graphicx}
\usepackage{dcolumn}
\usepackage{bm}
\usepackage[pdfpagelabels]{hyperref}

\def\be{\begin{equation}}
\def\ee{\end{equation}}
\def\bea{\begin{eqnarray}}
\def\eea{\end{eqnarray}}

\begin{document}

\title{Axionic Strings, Domain Walls and Baryons}

\author{Francesco Bigazzi$^{1}$, Aldo L. Cotrone$^{1,2}$, Andrea Olzi$^{1,2}$}
\affiliation{
$^1$ {INFN, Sezione di Firenze; Via G. Sansone 1; I-50019 Sesto Fiorentino (Firenze), Italy.\\}
$^2$ {Dipartimento di Fisica e Astronomia, Universit\'a di Firenze; Via G. Sansone 1; I-50019 Sesto Fiorentino (Firenze), Italy.\\}
}

\email{bigazzi@fi.infn.it, cotrone@fi.infn.it, andrea.olzi@unifi.it}

\begin{abstract}

When axionic strings carry a global charge, domain walls bounded by such strings may not be allowed to decay completely.
This happens in particular in some models where a composite 
axion-like particle is the pseudo-Nambu-Goldstone boson of chiral symmetry breaking of an extra quark flavor.
In this case, the global symmetry is the extra flavor baryonic symmetry.
The corresponding axionic domain walls can carry a baryonic charge: they represent the low energy description of the baryons made by the extra quark flavor.
Basic properties of these particles, such as spin, mass scale, and size are discussed.
The corresponding charged axionic strings are explicitly constructed in a specific calculable model.  

\end{abstract}

\maketitle

{\bf Introduction} - If the Peccei-Quinn (PQ) symmetry is broken after inflation, the abundance of axions depends on the decay pattern of axionic strings and domain walls (DWs) \cite{Kibble:1976sj,Vachaspati:1986cc,Chang:1998tb}. Axionic strings form at the Peccei-Quinn scale $f_a$, while domain walls, bounded by strings, are supposed to form around the confinement scale $\Lambda$. In the standard picture, when the anomaly coefficient $N_w$ is equal to one, the DWs ending on strings decay completely, mostly into axions and gravitational waves.

We consider a scenario where 
an axion-like particle (ALP) is the pseudo-Goldstone boson of the breaking of an axial $U(1)_A$ acting on just one extra massless quark flavor. In this case the condition $N_w=1$ is automatically realized as it happens in Kim-Shifman-Vainshtein-Zakharov-like axionic models. The extra flavor condenses at a scale $f_a \gg \Lambda$. This could be due e.g.~to quartic Nambu-Jona-Lasinio (NJL)-like interactions or some other mechanism preserving $N_w=1$. The basic feature of this scenario is the presence of the residual vectorial symmetry. We call it $U(1)_b$ because it is the extra flavor baryonic symmetry. 

One of the main points of this paper is that in such a scenario, some DWs could not decay completely, due to baryonic charge localized on their boundaries, i.e.~on the axionic strings. This is similar to what happens to charged strings (vortons) \cite{Carter:1993wu}. In fact, in such models, sufficiently small charged DWs can form from charged string loops before the confining transition. The charged DWs, which have a pancake shape and host a Chern-Simons theory on their world-volume \cite{Komargodski:2018odf}, describe at low energies the baryons composed by the extra quark flavor, so they are named ``axionic baryons'' or ``abaryons'' for short. These particles could constitute some fraction of dark matter, 
although we postpone the study of their phenomenology to the future.

The idea of stable remnant domain walls is not bound to their baryonic nature: it just depends on the existence of a conserved global charge under which the domain wall can be the lightest charged object. On the other hand, the scenario we actually investigate here is the one where the charge is actually associated to a baryonic symmetry.

In the first part of this paper, we introduce the charged DW description of these baryons and exhibit their main properties (mass scales, dimensions, etc.). 
An example of similar DWs, but still without the baryonic charge, has been recently explored in \cite{olzi}.
In the second part of the paper, as a first step towards the quantitative characterization of these baryons in a concrete model, we provide an explicit construction of the related axionic strings (see e.g.~\cite{Forbes:2000et} for related configurations). 
We calculate their tension and thickness above the confinement scale and provide their effective action.

{\bf Introducing Abaryons} - We consider models of composite QCD axions or axion-like particles. The axion is the Goldstone boson of the breaking of the axial $U(1)_A$ of an extra massless quark flavor condensing at a scale $f_a \gg \Lambda$, where $\Lambda$ is the dynamical scale of the $SU(N)$ Yang-Mills (YM) sector responsible for confinement. In the ALP case, the $SU(N)$ YM plus extra flavor system is a completely hidden sector, interacting only gravitationally with the Standard Model. In the QCD axion case, the $SU(N)$ YM is the standard color $SU(3)_c$. In both cases, the condensation of the extra flavor is due to some higher dimensional operator such as an NJL-type interaction term (see \cite{Bigazzi:2019eks} for a concrete realization). For many respects, the axion in these theories behaves as the $\eta'$ in single-flavored QCD. In the low energy limit, the Chiral Lagrangian of the latter particle has been argued to have a ``Hall droplet baryon" solution describing a single-flavor baryon \cite{Komargodski:2018odf}\footnote{While the considerations in \cite{Komargodski:2018odf,olzi} are performed in the planar limit, finite $N$ effects can modify the picture only quantitatively, not qualitatively.}. This is a pancake-shaped configuration formed by a gluonic core and an $\eta'$ profile interpolating from 0 to $2\pi$ as one crosses its world-volume, which comes with a non-trivial baryonic charge along its boundary. This is the same as a quantum Hall droplet: the charge is encoded in a chiral edge mode. In fact, the baryon hosts on its world-volume a $U(1)_N$ Chern-Simons (CS) theory. The baryon is stable, as it is the lightest object carrying one unit of baryonic charge.

The basic point is that a similar object exists for the composite axion models considered above \footnote{While this paper was in preparation, \cite{Ma:2019xtx} appeared with a comment on such a configuration.}.
In fact, since the axion comes from an extra quark flavor, there exists the $U(1)_b$ baryonic symmetry under which the ordinary axionic domain wall boundary (the string) can be charged. So, the baryon (``abaryon'') composed by the extra quark has a quantum Hall droplet description.

Already at the qualitative field theory level, one can argue that the spin of this baryon is $N/2$, where $N$ is the number of colors, precisely as in the Fractional Quantum Hall Effect (FQHE) \cite{Komargodski:2018odf} (that is, the spin of the abaryon is $3/2$ in the QCD axion case). The scaling behavior of a number of properties of the Hall droplet baryon was derived in \cite{Komargodski:2018odf}. In order to make progress 
one has to study the problem in an explicit model. In \cite{olzi} we have derived some properties of this system in a calculable example based on the so-called Witten-Sakai-Sugimoto (WSS) model. Its planar, strong coupling regime can be explored by means of the holographic correspondence which provides a dual classical gravitational description. The WSS model is the most fruitful top-down holographic model of QCD \cite{Witten:1998zw,Sakai:2004cn}. It shares the same vacuum structure with (planar) QCD.  Many observables of QCD can be calculated with ${\cal O}(10\%-20\%)$ accuracy. In this model, the scale of condensation $f_a$ is set by the strongly coupled version of a non-local quartic NJL interaction \cite{Antonyan:2006vw}.

In the dual gravitational description of the confined regime of the WSS model, the background is the one generated by $N$ D4-branes wrapped on a circle with anti-periodic boundary conditions for fermions. The corresponding ten-dimensional geometry includes four Minkowski directions, a four-sphere and a so-called cigar, i.e.~a two-dimensional manifold formed by a radial direction $u$ and a circle with coordinate $x_4$ which smoothly shrinks to zero size at the tip of the cigar, located at a certain position $u_0$. This parameter sets the dynamical scale $\Lambda$.
Type IIA strings on this background provide the gluonic sector of the theory.

\begin{figure}
\begin{center}
\scalebox{0.7}{\includegraphics{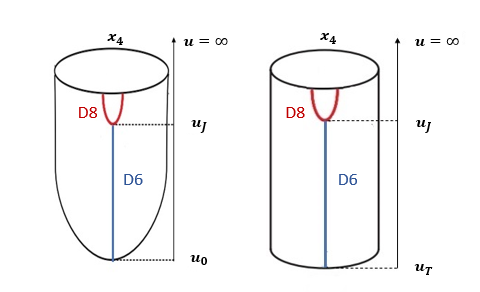}}
\caption{The embeddings of the D8-brane and D6-brane on the cigar (left) and on the cylinder (right), in the ALP case.}
\label{figsetup}
\end{center}
\end{figure}

The extra flavor degrees of freedom are introduced by means of a probe D8-brane embedded in the background. The codimension-one D8-brane describes a curve on the cigar, see figure \ref{figsetup} (left plot).
The position $u_J$ of the tip of the curve is related to the scale $f_a$. 
The D8-brane has a gauge field on its world-volume, whose modes are dual to extra flavor mesonic fields. The lowest such mode, which is the (pseudo-)Goldstone of chiral symmetry breaking, is the axion 
 \cite{Bigazzi:2019eks} \footnote{Gravitational wave emission in such a scenario has been studied in \cite{Bigazzi:2020avc}.}.

Let us first consider the ALP case.  In this setup, the abaryon has a ``gluonic core'' formed by a D6-brane, wrapped on the four-sphere, with a boundary on the D8-brane \cite{olzi} (see also \cite{Dubovsky:2011tu}).
With ``gluonic core'' we denote a description of the abaryon in terms of the explicit D6 brane: the flavor part (the D8-brane) is not essential for this description, making it evident that the degrees of freedom composing this object are purely gluonic, exactly as for the baryon vertex in the baryon. Since the thickness of the D6 is small, as we are going to show, this configuration gives the ``core'' of the abaryon. Analogously, its tension will be shown to be large, hence we are going to call it also ``hard core''. 

\begin{figure}
\begin{center}
\scalebox{0.5}{\includegraphics{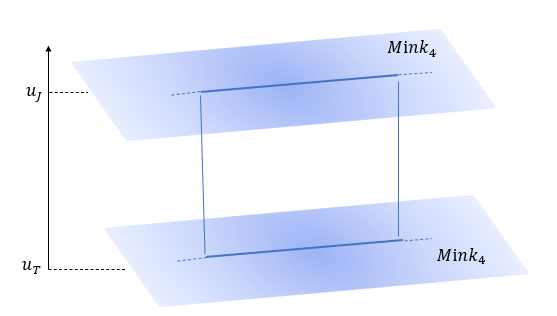}}
\caption{The D6-brane has a seven-dimensional world-volume, it is wrapped on a four-cycle and extended along $u$ and two Minkowski directions. Therefore, it has the shape of a string from the perspective of the Minkowski spacetime.}
\label{axionicstring}
\end{center}
\end{figure}

In the deconfined geometry (see figure \ref{figsetup}, right panel) the D6-brane attached to the D8-brane can terminate at the horizon placed at $u_T$. From the perspective of the Minkowski spacetime this embedding describes an axionic string (see figure \ref{axionicstring}). However, in the confined geometry (see figure \ref{figsetup}, left part), the part of the D6-brane not attached to the D8 has no place to terminate. Thus, either the DW has to be infinitely extended, or it can be extended between two strings (that is, the D6 has a second boundary on the D8 at a different position in Minkowski directions), or it must be completely bounded by a string - that is, the D6 has a single boundary on the D8 with the topology of a circle. This latter configuration, if charged, corresponds to the abaryon.

If the size of the abaryon is large, the D6-brane is expected to have a pot-like shape, with an almost ``vertical'' part (representing the stringy boundary of the abaryon) extending from the D8-brane at $u_J$ to (almost) the tip of the cigar at $u_0$, where the D6-brane is basically ``horizontal'' at constant $u_0$.
In \cite{olzi} we have analyzed the case where the tip of the D8-brane at $u_J$ corresponds to the tip of the cigar at $u_0$.
The properties of the DW, in this case, are a good approximation of the ones of the abaryon in the limit where the latter has a large extension.
With this caution remark in mind, we can take a number of information on the abaryon from \cite{olzi}.
The gluonic core has tension $T_{core} = \lambda^2 N \Lambda^3/3^6 \pi^3$ and thickness $\delta_{core} \sim (\lambda^{1/2}\Lambda)^{-1}$, where $\lambda$ is the 't Hooft coupling at the scale $\Lambda$.  
Moreover, its boundary attached to the D8-brane corresponds to the core of the axionic string.
Its tension and thickness are estimated in the next section, again in the large-size scenario.

The abaryon has also a ``mesonic shell'' description formed by the ALP and the other extra flavor meson profiles. We denote it as a ``shell'' because its thickness is much larger than the core's one. 
Since its tension is smaller than the one of the core, we are going to call it also ``soft shell''. It is more difficult to estimate the properties of this shell without an explicit solution, but the parametric scaling is very likely to be the same as in  \cite{Komargodski:2018odf,olzi}, that is $T_{shell} \sim \lambda^2 N^{1/2} \Lambda^3$ for the tension and $\delta_{shell} \sim N^{1/2}(\lambda\Lambda)^{-1}$ for the thickness.

\begin{figure}
\begin{center}
\scalebox{0.6}{\includegraphics{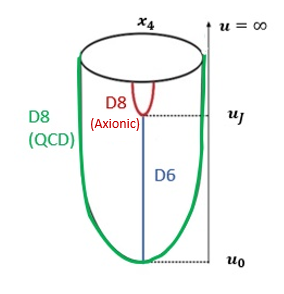}}
\caption{Branes' embeddings in the QCD axion case. The gauge sector is in the confined phase. The D6-brane can extend from the axionic D8-brane (red) to the D8-brane associated to the ordinary quarks (green).}
\label{QCDaxion}
\end{center}
\end{figure}

Let us now comment on the case of the QCD axion. In the dual gravitational background, there are other D8-branes supporting the ordinary Standard Model quark degrees of freedom, with smaller radial position of the tip. Thus, there are two types of D8-branes: the one associated to the axion we have discussed so far (and depicted in figure \ref{figsetup}), which we may call the ``axionic D8-brane'', and the D8-branes associated to the Standard Model flavors, which we may call the ``ordinary quark  D8-branes''. In this configuration (see figure \ref{QCDaxion}) the D6-brane can also extend from the ``axionic D8-brane'', where it has a boundary, down to the ``ordinary quark D8-branes'', where it has a second boundary. This configuration can have a cylindrical shape, without the ``horizontal'' portion of the D6-brane. If this case is realized instead of the one described above, the DW does not have a hard core world-volume with a CS theory, but just a soft mesonic shell (mostly axionic and $\eta'$) \cite{Gabadadze:2000vw}, and it is basically a vorton with both $U(1)_b$ and ordinary baryon number. 

In fact, depending on the parameter ratio $T_c/T_a$, the dominant configuration at a given temperature could be the one with the two boundaries just described (surely dominant at large $T$) or the abaryon configuration with a single boundary at $u_J$ as in the ALP case (if $T$ is sufficiently smaller than $T_a$). 
The main difference with respect to the ALP case is that now there are the ordinary quark D8-branes localized at $u_0$ which can host an ordinary baryon. Therefore in the QCD axion case, the abaryon could decay to ordinary baryons, so it would be metastable.

An interesting observable is the amount of energy that is stored in these particles at equilibrium.
In order to give a precise value of the dimension and mass of the abaryon, one has to construct the explicit solution. 
We defer this endeavor to a future study.
In the next section we construct the axionic string in the deconfined phase, which determines the local physics of the boundary of the abaryon (where axions are mostly emitted), and of course, is of interest on its own. The model at hand has a rich zoology of strings and walls, which will be presented elsewhere.

{\bf Axionic String} - When the temperature $T$ is larger than the critical temperature for deconfinement, which in the WSS model is $T_c= \Lambda/(2\pi)$, the theory is in the state described holographically by the ten-dimensional background \cite{Aharony:2006da}
\begin{eqnarray}
&& ds^{2} =\left(\frac{u}{R}\right)^{3/2} \left[-f_{T}(u)dt^{2} + dx^i dx^i + f_{T}(u)d x_4^2\right] \nonumber \\
&& \quad \qquad +\left(\frac{R}{u}\right)^{3/2}\left[\frac{du^{2}}{f_{T}(u)} + u^{2}d\Omega_4^2\right]\,, \\
&& e^{\Phi} = g_{s}\left(\frac{u}{R}\right)^{3/4}\,, \quad F_{4} = \frac{3R^{3}}{g_{s}}\omega_4\,, 
\end{eqnarray}
with 
$f_{T}(u) = 1 - u_{T}^{3}/u^{3},\ R^{3} = \pi g_{s}Nl_{s}^{3}$,
where $\Phi, F_4$ are the dilaton and Ramond-Ramond four-form field strength, $u_T$ is the radial position of the horizon (related to the temperature through $9\,u_T = 16\pi^2 R^3 T^2$), the $(x_4, u)$ subspace forms a cylinder (replacing the cigar of the confined phase), $\omega_4$ is the volume form of the four-sphere $\Omega_4$ and $g_s, l_s$ are the string coupling and length.
The D8-brane has a profile on the cylinder, which is represented in figure \ref{figsetup} (right plot), whose embedding is described by the equations that can be found e.g.~in \cite{Aharony:2006da,Bigazzi:2019eks}. 

The axionic string has a ``hard core'' holographically described as a D6-brane.
The latter is wrapped on the four-sphere, extended in two Minkowski directions which we will call $t, x_1$ and extended in the holographic radial direction from the tip of the D8-brane at $u_J$ to the horizon at $u_T$, as depicted in figures \ref{figsetup} and \ref{axionicstring}.
In the regime $T_c < T < T_a$, where $T_a$ is the temperature of the PQ transition \footnote{In this model $T_a \sim \left(16 \pi^3 (0.153)^2 \Lambda f_a^2 / \lambda \right)^{1/3}$ \cite{Bigazzi:2019eks}.}, this is true both in the ALP and QCD axion cases, since the ordinary quark D8-branes are in a chiral symmetry restored configuration, so the D6-brane does not intersect them. 
The tension $T_{hard}$ of this hard core is readily calculated from the action of the D6-brane (with tension $T_6 = (2\pi)^{-6}l_s^{-7}$)
\be
S=-T_6\int d^7x \,e^{-\Phi}\sqrt{-g_7}\equiv T_{hard} \int dt dx_1\,,
\ee
where $g_7$ is the determinant of the induced metric on the D6-brane. 
In the $u_J\gg u_T$ limit, the result (using formulas in appendix B of \cite{Bigazzi:2019eks}) is
\be\label{Thard}
T_{hard} \simeq \frac{1}{3^5 \pi}\lambda^2 N T_c^2 \left[\frac{3^4}{2^8}\frac{\widetilde{T}_a^4}{c_a^4} - \widetilde{T}^4 \right]\,,
\ee
where $c_a \sim (0.1538/0.7)$ is a coefficient of the model
and
$\widetilde{T}_a = T_a / T_c\,, \widetilde{T}= T / T_c$. The tension grows as the temperature drops towards $T_c$ and reaches its maximum in the supercooling regime $\widetilde T \rightarrow 0$. The expression of $T_{hard}$ is the same for a circular string of radius $L$, whose total core energy is thus $2\pi L T_{hard}$ \footnote{These formulas are strictly valid for $u_J \gtrsim 10 u_T$, which corresponds to the phenomenological sensitive regime $T_a \gg T_c$ (or $f_a \gg T_c$), but it can be shown, using the formulas in \cite{Bigazzi:2019eks}, that for $u_J < 10 u_T$ the discrepancy with the actual results is below $5\%$.}.
The thickness $\delta_{hard}$ of this core is provided by the brane thickness, which is roughly the local string length scale and has its largest value at $u_T$, giving
\be 
\delta_{hard} \sim \left(\sqrt{\lambda \widetilde{T}^3}\,T_c\right)^{-1}\,. \label{dhard}
\ee
Note that the (inverse) thickness is naturally set by the temperature scale $T$, rather than the scale of symmetry breaking $T_a$ as commonly assumed in the literature for axionic strings. This is because the transverse size of the D6-brane, as seen by the theory on the boundary (where it represents the axionic string), depends on the holographic direction $u$. Objects localized at $u_J$ have a smaller size than objects localized at $u_T$. Since the string is really a brane stretched from $u_J$ to $u_T$ (see figure \ref{figsetup}, right panel, and figure \ref{axionicstring}), its maximal transverse size is set at $u_T$, giving formula (\ref{dhard}). Only its portion close to the D8-brane would have size set by $u_J$, giving a thickness set by the symmetry breaking scale as $\delta \sim \left(\sqrt{\lambda \widetilde{T}_a^3}\,T_c\right)^{-1}$.

A completely analogous computation can be performed in the confined phase, for the QCD axion strings with an axionic and an $\eta'$ profile (D6-brane between the ``axionic'' and ``regular quark'' D8-branes), and for the boundary of the abaryon in the ALP case if its dimension is large (D6-brane with pot-like shape). In the standard setup where the ``regular quark" D8-branes have their tip in $u_0$, the tension and thickness of the core of the strings are calculated as above but for $u_T \rightarrow u_0$, with the results
\begin{eqnarray} 
&& T_{hard} \sim \frac{1}{2^2 3^5 \pi^3}\lambda^2 N \Lambda^2 \left[ \left(\frac{d_a f_a^2}{\lambda N \Lambda^2}\right)^{4/3} - 1 \right]\,, \\
&& \delta_{hard} \sim \left(\sqrt{\lambda} \Lambda \right)^{-1}\,,\quad d_a \sim (2(2.4)3^3  \pi)\,,
\end{eqnarray}
where the tension is given in the $u_J\gg u_0$ limit.
Note that in the WSS model $f_a^2$ scales as $\lambda N \Lambda^2$.

Coming back to the deconfined phase, the axionic string can be also seen as having a ``soft mesonic (axionic) profile'', described holographically by the Abelian gauge field  $A$ on the D8-brane. The D8-brane world-volume also hosts a scalar mode corresponding to its fluctuations in the transverse direction. In principle, this mode could be turned on too (in fact, it is considered in \cite{olzi}). However, in the quadratic order low-energy approximation we are going to employ, the scalar mode is decoupled from the gauge field, so for simplicity, we are going to keep it off. Of course, it would be interesting to consider its contribution to the energy, which is expected to be similar to the gauge field one, and crucial to describe the full non-linear BIon configuration along the lines of \cite{Callan:1997kz}.

Upon reduction on the four-sphere, the theory of $A$ at low energy is simply Maxwell-Chern-Simons in five curved dimensions. In order to analyze the latter theory, it is convenient to parameterize the radial coordinate as $u(z) = u_{J}k^{1/3}(z),\ k(z) = 1+z^{2}$, such that the tip of the D8-brane $u_J$ is located at $z=0$. The axion is the $z$-integral of the component $A_z$ of the gauge field \cite{Bigazzi:2019eks}. Considering the symmetries of the configuration, it is sufficient to employ the ansatz with only three field strength components turned on, $F_{x_2 z}, F_{x_3 z}, F_{x_2 x_3}$. The action for these components of the field strength can be derived by expanding the D8-brane action to quadratic order, obtaining
\begin{eqnarray}\label{action}\nonumber
S = -\frac{\lambda N}{3^2 2^6 \pi^4}\frac{\widetilde{T}_a}{c_a}\int d^{4}x\,dz\Bigl[\sqrt{\frac{z^2\,f_{T}(z)}{k^{5/3}(z)\gamma_{T}(z)}}F_{x_2 x_3}^{2}+ \\ 
\frac{9T_a^2}{4 c_a^2}\sqrt{\frac{k^3(z)f_{T}(z)\gamma_{T}(z)}{z^2}}\left(F_{x_2z}^{2}+ F_{x_3z}^{2}\right)\Bigr],\,\,\,\,\,\,
\end{eqnarray}
where
$\gamma_{T}(z) = (u^{8}(z)f_{T}(z) - u^{8}_{J}f_{T}(0))/u^{8}(z)$.
The equations of motion can be solved in terms of an unknown function $H_T(x_2,x_3,z)$ as
\begin{eqnarray}
&& F_{x_{3}z} =  \frac{|z|}{k^{3/2}(z)}\frac{\partial_{x_{2}}H_{T}}{\sqrt{f_{T}(z)\gamma_{T}(z)}}\,,\\
&& F_{x_2z} = -\frac{|z|}{k^{3/2}(z)}\frac{\partial_{x_3}H_{T}}{\sqrt{f_{T}(z)\gamma_{T}(z)}}\,,\\
&& F_{x_2x_3} = \frac{9T_a^2}{4 c_a^2}\sqrt{\frac{k^{5/3}(z)\gamma_{T}(z)}{z^2 f_{T}(z)}}\,\partial_{z}H_{T}.\,\,
\end{eqnarray}

The part of the D6-brane attached to the D8-brane sources the gauge field.
It represents a linear distribution of magnetic charge extended along $x_1$.
Thus, it gives a source term in the Bianchi identity
\be
dF = -2\pi\sqrt{2}\,\delta(x_{2})\delta(x_{3})\delta(z)dx_{2}\wedge dx_{3}\wedge dz\,.
\ee
This corresponds to an equation for $H_T$
\begin{eqnarray}
&&\,\,\,\,\,\,\,\,\frac{9T_a^2}{4 c_a^2}\partial_{z}\Bigl[\sqrt{\frac{k^{5/3}(z)\gamma_{T}(z)}{z^2 f_{T}(z)}}\,\partial_{z}H_{T}\Bigr] + \nonumber\\
 &&\frac{|z|\left(\partial_{x_2}^{2} + \partial_{x_3}^{2}\right)H_{T}}{\sqrt{k^3(z)f_{T}(z)\gamma_{T}(z)}} = -2\pi\sqrt{2}\,\delta(x_{2})\delta(x_{3})\delta(z)\,.\,\,\,\,\,\,\,\,\,\,\,\,\,\,
\end{eqnarray}
We look for a series expanded solution of the kind
\begin{equation}
H_{T} = \sum_{n=0}^{\infty}\zeta_{T,n}(0)\zeta_{T,n}(z)Y_{T,n}(r)\,, \quad r=\sqrt{x_2^2+x_3^2}\,,
\end{equation}
where we have set to zero the collective coordinates of the axionic string (see \cite{olzi}).
The modes $Y_{T,n}$ correspond to four-dimensional mesons at finite temperature.
The Bianchi identities are solved by the two relations
\begin{eqnarray}\label{eigeneq}
&& -\partial_{z}\left[\sqrt{\frac{k^{5/3}(z)\gamma_{T}(z)}{z^2 f_{T}(z)}}\,\partial_{z}\zeta_{T,n}(z)\right] = \\ \nonumber
&&= \frac{|z|}{k^{3/2}(z)}\frac{\lambda_{T,n}\zeta_{T,n}(z)}{\sqrt{f_{T}(z)\gamma_{T}(z)}}\,,\\ 
\label{eqforY} && \left(\partial_{x_2}^{2} + \partial_{x_3}^{2} - \frac{9\widetilde{T}_a^2 T_c^2}{4 c_a^2}\lambda_{T,n}\right)Y_{T,n}(r) =\\ 
&& = -2\pi\sqrt{2}\,\delta(x_{2})\delta(x_{3})\,, \nonumber
\end{eqnarray}
which lead to the completeness relation for the eigenfunctions $\zeta_{T,n}(z)$
\begin{equation}
\sum_{n=0}^{\infty}\frac{|z|}{k^{3/2}(z)}\frac{\zeta_{T,n}(0)\zeta_{T,n}(z)}{\sqrt{f_{T}(z)\gamma_{T}(z)}} = \delta(z)\,.
\end{equation}

One can integrate numerically equation (\ref{eigeneq}) in order to extract the eigenvalues $\lambda_{T,n}$, by requiring that the derivatives of the functions $\zeta_{T,n}(z)$ vanish at $z\to\pm\infty$.
Then, equation (\ref{eqforY}) is solved in terms of the modified Bessel function of order zero of the second kind $Y_{T,n}(r) = \sqrt{2} K_0(\frac{3T_a}{2 c_a}\sqrt{\lambda_{T,n}}r)$, apart from the zero mode corresponding to zero eigenvalue, which is a simple logarithm $Y_{T,0}(r) = -\sqrt{2} \log{(T_a r)}$; the latter corresponds to the Goldstone boson of the spontaneous chiral symmetry breaking, i.e. the ALP of the theory. If $\tilde{b}\equiv u_T/u_J\ll1$, the eigenvalues can be well approximated by their values for $\tilde{b}=0$. The ones corresponding to the first even eigenfunctions (odd modes do not enter in $H_T$) are $\lambda_{T,n} = 0, 2.12, 6.23, 12.35, 20.51, ...$ Imposing the orthonormality condition
\begin{equation}
\int dz\,\frac{|z|}{k^{3/2}(z)}\frac{\zeta_{T,n}(z)\zeta_{T,m}(z)}{\sqrt{f_{T}(z)\gamma_{T}(z)}} = \delta_{n,m}
\end{equation}
for the functions $\zeta_{T,n}$ and plugging the ansatz for the field strength into the action (\ref{action}), we get for the axionic string modes
\begin{equation}
\widetilde{Y}_{T,n}(r) \equiv \sqrt{\frac{N\lambda\widetilde{T}_a^3}{2^{7} \pi^4 c_a^3}}\,T_c\, \zeta_{T,n}(0)Y_{T,n}(r)\,,
\end{equation}
the four-dimensional effective action
\begin{equation}\label{effaction}
S = -\frac{1}{2}\sum_{n=0}^{\infty}\int d^{4}x\left[\left(\partial_{r}\widetilde{Y}_{T,n}(r)\right)^2 + \frac{9T_a^2 \lambda_{T,n}}{4 c_a^2}\widetilde{Y}_{T,n}^{2}(r)\right]\,.
\end{equation}
Note that this global string effective action is rigorously derived from the fundamental theory - it is not postulated.
It can be seen as the global string analog of the Chiral Lagrangian coupled to an infinite tower of massive mesonic modes. 

We can extract the tension of the soft part of the axionic string from the action evaluated on-shell on the solution
\begin{eqnarray} \label{tensionint}
T_{soft} &=& \frac{1}{2^{6} \pi^3 c_a^3}\lambda N T_c^2 \widetilde{T}_a^3 \int_0^{\infty} dr \Bigl\{\zeta_{T,0}^2(0)\frac{1}{r} + \\
&& 
+\, r \frac{9T_a^2}{4 c_a^2}\sum_{n=1}^{\infty} \zeta^2_{T,n}(0) \lambda_{T,n}  \left[K_0^2 + K_1^2 \right] \Bigr\}\,,\nonumber
\end{eqnarray}
where $K_1$ is the modified Bessel function of order one and the argument $(\frac{3T_a}{2 c_a}\sqrt{\lambda_{T,n}}r)$ of $K_0$ and $K_1$ is understood. The divergence of the tension at the location of the string ($r=0$) should be cut off by the hard core scale $\delta_{hard}$ in (\ref{dhard}). In fact, the quadratic effective theory on the world-volume of the D8-brane is not suitable to describe the string core. For example, the field strength diverges in that region, so that the employed approximation breaks down. But the brane construction gives a precise description of the string core: it is the boundary of the D6-brane.
It has a gluonic nature and has a thickness given in (\ref{dhard}), which is thus the natural cut-off in the effective ``soft'' description \footnote{In a BIon-like description along the lines of \cite{Callan:1997kz}, where the D6 would be seen as a ``spiky'' portion of the D8 protruding in the $u$ direction, the divergence at $r \sim 0$ would be due to the infinite length of the spike. But in the present case the spike would not be infinite: it would end at $u_T$, precisely corresponding to a size $r \sim \delta_{hard}$ (remember that $\delta_{hard}$ is the thickness evaluated at $u_T$).}.

The large distance divergence in (\ref{tensionint}) is the usual one of cosmic global strings, so that
\be \label{logdiv}
T_{soft} \sim \lambda N T_c^2  \log{(R_c /\delta_{hard})}\,,
\ee
$R_c$ being the large distance cut-off which is, assuming that there is one cosmic string per Hubble patch, the inverse Hubble scale.

{\bf Charged Mode Solution} - The axionic string can have a $U(1)_b$ charged mode, described again by the D8-brane gauge field. 
In this section, we derive an approximate solution for the gauge field components $A_t$ (associated with the baryon number) and $A_{x_{1}}$, that does not change the monopole-like solution found in the previous section, following the same kind of approximation employed for the baryons in \cite{Hashimoto:2008}. First, we find a solution very close to the source, where we can neglect the curvature of spacetime. In this region, we are able to solve the full equations of motion derived from the five-dimensional Maxwell-Chern-Simons action. As we will show, the solution is very localized around the string position in the large $\lambda$ limit. As we move away from the string, we can neglect the Chern-Simons term since it is sub-leading in the $1/\lambda$ expansion. The solution in this intermediate (again flat) region matches with the expansion of the full flat-space solution far from the source. Finally, we connect this solution to the large-$z$ region of five-dimensional curved space-time, where we solve the linearized equations of motion.

In the flat space limit, where $k(z) = 1$, we employ the ansatz $A_{t}(\rho), A_{x_1}(\rho)$, where $\rho = \sqrt{x_2^2 + x_3^2 +w^2}$, the coordinate $w$ being related to $z$ through the rescaling $w(z) = \sqrt{\frac{4}{3(8 - 5\tilde{b}^3)}}\frac{c_a}{T_a}z$. 
The equations of motion, again derived from the expansion of the D8-brane action, read
\begin{eqnarray} \label{eqflat1}
&\partial_\rho(\rho^2\partial_\rho A_t) - \mu\sqrt{1 - \tilde{b}^3}\partial_\rho A_{x_1} = 0\,,\\
&\partial_\rho(\rho^2\partial_\rho A_{x_1}) - \frac{\mu}{\sqrt{1 - \tilde{b}^3}}\partial_\rho A_t = 0\,,\label{eqflat2}
\end{eqnarray} 
where $\mu= (3\sqrt{2}\pi^2 c_a^2 T_c)/(\lambda T_a^2)$.
They can be integrated analytically, neglecting the divergent solution in $\rho = 0$ and a constant term, to get
\begin{eqnarray}\label{solflat1}
    &A_t(\rho) = c\,\sqrt{1 - \tilde{b}^3}\,e^{-\mu/\rho}\,,\\
    &A_{x_1}(\rho)= c\,e^{-\mu/\rho}\,,\label{solflat2}
\end{eqnarray}
with an arbitrary constant $c$. Since $\mu\sim 1/\lambda \ll 1$, the maximum of the corresponding field strength is much greater than one, and it occurs very close to $\rho = 0$.

We can employ a $1/\lambda$ expansion and look for a solution in this limit, called the scaling limit. Since, as we have just seen, the solution is very localized around $\rho=0$,
in the (still flat) intermediate region $1/\lambda \ll \rho \ll 1$ the Chern-Simons term (the one multiplied by $\mu$ in (\ref{eqflat1}), (\ref{eqflat2})) is sub-leading and does not enter the equations motion for $A_{t,x_1}$, which have solutions in terms of the flat three-dimensional Green function
\begin{equation}\label{intermediate}
    A_{t,x_1}(\rho) \sim \frac{1}{\rho},
\end{equation}
with different proportionality constants for the two modes. These solutions are indeed the exponential ones (\ref{solflat1}), (\ref{solflat2}) for $\rho \gg \mu$, up to a constant. 

Now we look at large distances, in order to find the solution in the curved background. 
In this region the ansatz is slightly modified to $A_t(r,z)$, $A_{x_1}(r,z)$, $r=\sqrt{x_2^2+x_3^2}$.  
As it is clear from formula (\ref{intermediate}) in the flat space limit, the solutions become smaller as one moves away from the source. 
Thus, it makes sense to linearize the equations of motion at large distances (considering also the suppression with $\lambda$ of the CS term).
A-posteriori, one can check that indeed the solutions at large distances are actually small.
The linearized equations read
\begin{eqnarray*}
    &\frac{|z|}{k^{5/6}}\frac{1}{\sqrt{f_T\gamma_T}}
    \frac{1}{r} \partial_r ( r \partial_r A_t) 
    + \frac{9T_a^2}{4c_a^2}\partial_z\left(\frac{k^{3/2}}{|z|}\sqrt{\frac{\gamma_T}{f_T}}\partial_z A_t\right)=0,\\
    &\frac{|z|}{k^{5/6}}\sqrt{\frac{f_T}{\gamma_T}}
    \frac{1}{r} \partial_r ( r \partial_r A_{x_1}) 
    + \frac{9T_a^2}{4c_a^2}\partial_z\left(\frac{k^{3/2}}{|z|}\sqrt{f_T\gamma_T}\partial_z A_{x_1}\right)=0.
\end{eqnarray*}
The solutions can  be written in terms of a series expansion
\begin{eqnarray}
    &&A_t(r,z) = \sum\limits_{n=1}^{\infty}\alpha_n(0)\alpha_n(z)Y_n(r),\\
    &&A_{x_1}(r,z) = \sum\limits_{n=1}^{\infty}\beta_n(0)\beta_n(z)Y_n(r),
\end{eqnarray}
where $\alpha_n$ and $\beta_n$ are two sets of eigenfunctions defined by
\begin{eqnarray*}\label{alphaeq}
    &&\partial_z\left(\frac{k^{3/2}}{|z|}\sqrt{\frac{\gamma_T}{f_T}}\partial_z \alpha_n(z)\right) + a_n\frac{|z|}{k^{5/6}}\frac{\alpha_n(z)}{\sqrt{f_T\gamma_T}} = 0,\\
    &&\partial_z\left(\frac{k^{3/2}}{|z|}\sqrt{\gamma_Tf_T}\partial_z \beta_n(z)\right) + b_n\frac{|z|}{k^{5/6}}\sqrt{\frac{f_T}{\gamma_T}}\beta_n(z) = 0.
\end{eqnarray*}
They are the meson eigenfunctions at finite temperature and are normalized as
\begin{eqnarray*}\label{alphanorm}
    &&\int dz\,\frac{|z|}{k^{5/6}}\frac{\alpha_n(z)\alpha_m(z)}{\sqrt{f_T\gamma_T}} = \delta_{nm},\\
    &&\int dz\, \frac{|z|}{k^{5/6}}\sqrt{\frac{f_T}{\gamma_T}}\beta_n(z)\beta_m(z) = \delta_{nm}.
\end{eqnarray*}
The functions $Y_n$ are again the Bessel functions $K_0$ with masses $a_n$ for $A_t$, and $b_n$ for $A_{x_1}$. 
They are exponentially small at large $r$.
The equations of motion are homogeneous so the solutions are defined up to an overall constant. 
Thus, the charge and current densities in the dual theory at the boundary are arbitrary parameters in the straight string configuration.

{\bf Conclusions} - In this paper we have considered string and domain wall configurations in a (single-flavor) QCD-like theory where chiral symmetry breaking occurs in the deconfined phase.
This generates an axion-like particle - the pseudo-Nambu-Goldstone boson of the spontaneous axial symmetry breaking - and associated axionic strings.
The strings can carry a global charge, which is the baryonic charge of the condensing flavor.

We have explicitly constructed the straight (possibly charged) string solution in a strongly coupled holographic model, calculating its tension, thickness, and effective action. It is basically a global string model, exhibiting for example the usual logarithmic divergence in the tension (\ref{logdiv}). Its advantage, as compared to purely effective string models, is that we have access to the microscopic theory. This allows, for example, to understand the origin of the near-string divergence: it is due to the presence of a hard core which is not taken into account by the effective (soft) mesonic modes, and which is precisely described in the holographic model by a D6-brane boundary, i.e.~a gluonic field configuration. This hard core is the exact global string analog of the baryon vertex in the ordinary baryon. Moreover, the string modes in the effective action (\ref{effaction}) have a clear microscopic origin: they correspond to mesonic modes. Finally, the results of the computations give very precise values and behaviors in terms of the theory parameters, which provide novel useful information (e.g. see the comments around formulas (\ref{Thard}), (\ref{dhard})).
     
In the confined phase, the axionic strings must bound domain walls.
A crucial observation is that both string loops (in the deconfined phase) and disk-shaped domain walls bounded by string loops (in the confined phase) can be stable if they carry some units of charge.
Being the latter the baryon number, such charged domain walls in the confined phase are nothing else than single-flavor baryons - a novel observation as far as we are aware.
We have described some features of these baryons in the strongly coupled holographic example mentioned above.

Apart from the purely theoretical interest of the analyzed configurations, one could consider such QCD-like theories as dark sector candidates.
In fact, the scenario we have considered provides an axion-like particle (or even a composite QCD axion if the extra flavor is charged under our visible $SU(3)_c$ color symmetry) and (dark) baryons, as described above.
The corresponding cosmological scenario would entail a QCD-like dark sector which, cooling down in the expansion of the Universe, goes through two separate phase transitions. The first one is the chiral transition mentioned above, producing ALPs and (possibly charged) strings in the deconfined phase. 
The second transition is the confining one, producing uncharged DWs, baryons (i.e.~charged domain walls), and glueballs. These particles could compose (at least a fraction of) dark matter. 

In order to assess the phenomenological relevance of this scenario, explicit string loop and disk-shaped configurations must be constructed to extract their equilibrium properties, most importantly their mass. Besides these remnants, the string-wall network would eventually decay into axions producing a cosmologically relevant relic density. Moreover, the decay of topological defects can generate a stochastic background of gravitational waves that can potentially be observed in current and future experiments. We plan to report on such computations in the future.

{\bf Acknowledgments} - We thank Alessio Caddeo and Michele Redi for very helpful discussions.
ALC would like to thank Alessandro and David, for their interest hugely motivated this study.

\end{document}